\documentclass[12pt]{article}
\input{epsfig.sty}
\newcommand{\beq}{\begin{eqnarray}}
\newcommand{\eeq}{\end{eqnarray}}
\begin{document}
\title{QCD: The $\Lambda(1405)$ as a Hybrid}
\author{Leonard S. Kisslinger$^1$ and Ernest M. Henley$^2$\\
$^1$ Department of Physics, Carnegie Mellon University, Pittsburgh, PA 15213\\
$^2$ Department of Physics, University of Washington, Seattle, WA 98195}

\date{}

\maketitle

\begin{abstract} Using the QCD Sum Rule Method, we estimate the mass of the
lowest strange mixed hybrid/three-quark baryon with $IJ^P=0(1/2)^-$. We find 
the mass for a hybrid is approximately that of the $\Lambda(1405)$, whose 
nature has been a puzzle for many decades. Possible tests of this result are 
discussed.

\end{abstract} 

\section{Introduction}

   The nature of the $\Lambda(1405)$ has been of interest for many years.
More than four decades ago experiments showed\cite{efk65} that this  
was a state with spin=1/2. Using a SU(3) meson-baryon potential
it was predicted to be a $\bar{K}-N$ resonance\cite{dalitz61,dwr67} with 
$IJ^P=0(1/2)^-$ . There have been many studies using such a model. Recently 
chiral SU(3) meson-baryon potentials have been used in detailed 
studies\cite{oset09, weise08}, and conclude that the the $\Lambda(1405)$ is 
a $\bar{K}-N$ resonance. See Refs.\cite{oset09,weise08} for references to 
earlier publications using such a model. Recent experiments using the CLAS 
detector system at JLAB have carried out high statistics photoproduction 
measurements of the $\Lambda(1405)$\cite{rs09}. Although the analysis of the 
data is not complete, the present results do not seem to be consistent with 
$\bar{K}-N$ resonance models.

Recently, the method of QCD Sum Rules
has been used to explore the possibility that the $\Lambda(1405)$ has a
large pentaquark component\cite{oka08}. As explained in Ref\cite{oka08},
the results are not conclusive; and the authors also discuss attempts by 
other authors to find the nature of the $\Lambda(1405)$. 
These theoretical models and experiments have shown that
the $\Lambda(1405)$ is not a standard three-quark baryon, and are the 
motivation for our present work, in which we explore the possibility that
the $\Lambda(1405)$ is a strange hybrid baryon.

   For many decades there has been a great interest in
detecting and studying hybrid hadrons. The early work on hybrid hadrons
was based on quark models, in which mesons with certain so-called exotic 
quantum numbers could not exist as quark-antiquark states. Since with
three quarks there are no such exotic quantum numbers, all of the work
was on mesons as quark-antiquark-gluonic states. See Ref\cite{close88} for 
a review of hybrid mesons using bag models. 

   At the present time by hybrids one uses the concepts of Quantum 
Chromodynamics (QCD) in which the quarks are in a color octet, and with a
 color octet valence gluon form a color singlet physical state. Therefore, 
hybrids can be mesons or baryons.
With such a hybrid theory the $\Lambda(1405)$
would have three quarks (uds) in an octet configuration and a valence octet
gluon. Thus it would have a [udsg] structure. There have  been a number of 
studies of the $\Lambda(1405)$ as a strange hybrid. In recent studies by 
Kittel and Farrar\cite{kittel00,kittel05} using the bag model of Barnes and 
Close (who also studied the $\Lambda(1405)$\cite{barnes83a,barnes83b}) the 
isosinglet $\Lambda(1405)$ and $\Lambda(1520)$ (with $J^P = (3/2)^-$) were 
found to be hybrids. This model, however, is not consistent with the negative 
parity of the $\Lambda(1405)$. Note that in the earlier work with this 
model\cite{barnes83b} the lightest possible hybrid was found to be $P11(1710)$,
and a mass of 1405 was ruled out.

   Soon after the method of QCD Sum Rules was introduced it was used to
study hybrid mesons\cite{balitsky82,govaerts83}. Recently a study of vector 
$J^P=1^{--}$ states found\cite{guo07} $q\bar{q}g,\;q\bar{s}g,\;{\rm and\;}
s\bar{s}g$ hybrid mesons with masses 2.3-2.6 GeV. Some years ago a 
calculation using QCD Sum Rules was carried out to find
lightest hybrid baryon with $J^P=(1/2)^+$, like the proton\cite{marty91}.
This was followed by an improved calculation\cite{kl95}, with
the result that the mass of this hybrid is approximately that of the 
$P11(1440)$, the Roper resonance. The Roper is a very broad resonance, and
there might even be two states, a hybrid and a non hybrid. An extension of
this work, that is closely related to our present project, was the use
of QCD Sum Rules to find the lowest $J^P=(1/2)^+$ strange hybrid 
baryon\cite{lsk04}. Such a state was found at the energy of the 
$\Lambda(1600)$, about 500 MeV above the $\Lambda$ ,
which allows a possible experimental test via $\sigma$ decay. We
shall discuss this below.

Recently studies
of heavy quark hybrids were carried out. It was shown that there is no
satisfactory solution for hybrid charmonium or upsilon in the lowest
energy states\cite{kpr09}. In a subsequent study\cite{lsk09} it was shown 
that the $\Psi'(2S)$ and $\Upsilon(3S)$ states are approximately 50-50 
mixtures of hybrid and normal mesons. This provides a solution to some puzzles
in the decays of these states that standard quark models, which can fit
the energies of the states as pure heavy quark mesons, cannot explain.

   In the present research we extend the work of Ref\cite{kl95}, first using
a current for a $IJ^P=0(1/2)^-$ strange hybrid baryon, an then a current
with  mixed hybrid and  normal three-quark strange components.  Our  objective
is to find the mass of the lowest $IJ^P=0(1/2)^-$ mixed strange 
hybrid/three-quark baryon, to see if our solution satisfies the conditions 
for a satisfactory QCD Sum Rule solution, and to compare it to the mass of 
the $\Lambda(1405)$. We then discuss some possible experimental tests of 
the $\Lambda(1405)$ as a hybrid, incluidng photoproduction.

\section{$\Lambda(1405)$ as a hybrid}

  Following the work of Ref.\cite{kl95}, in which QCD Sum Rules were used
to show that there is a hybrid baryon with the quantum numbers of the
proton and the Roper resonance, with a mass about that of the
Roper resonance, we investigate the possibiity that the $\Lambda(1405)$
is a hybrid baryon. Also, recently it has been shown that although
the $\psi'(2S)$ is not a hybrid meson, it is a mixed hybrid-charmonium
meson, with a similar result for the $\Upsilon(3S)$, which solves some
puzzles\cite{lsk09}. In our present work we attempt to find the
lightest hybrid with the properties of the $\Lambda(1405)$; and if it turns
out that there is no satisfactory solution, in future work we shall explore 
the possibility that the $\Lambda(1405)$ is a mixed hybrid-normal baryon.

  The current that we use for a strange hybrid $I=0,J^P = 1/2^-$ baryon is 
that used in Ref.\cite{kl95} with modifications for the quantum numbers of
the $\Lambda(1405)$. The current for the $I=0,J^P = 1/2^+$ $\Lambda$ 
baryon\cite{cdks81}, expressed in a more convenient form\cite{rry85}
\beq
\label{1}
   J_{\Lambda}(x) &=& \sqrt{\frac{1}{2}} \epsilon^{abc}[(u^a(x) C\gamma^\mu 
s^b(x)) \gamma^5 \gamma^\mu d^c(x)-(d^a(x) C\gamma^\mu s^b(x)) \gamma^5
\gamma^\mu u^c(x)]\; .
\eeq
Using the same modification as for a hybrid nucleon\cite{kl95}, the current
that we use for a strange $I=0,J^P = 1/2^-$ hybrid is
\beq
\label{2}
J_H(x)& =& \sqrt{\frac{1}{2}} \epsilon^{abc}[(u^a(x) C\gamma^\mu 
s^b(x)) \gamma^\alpha [G_{\mu \alpha} d(x]^c-(d^a(x) C\gamma^\mu s^b(x))
 \gamma^\alpha [G_{\mu \alpha} u(x]^c] \; .
\eeq

In Eq(\ref{2}) a,b,c are color indices and u, d, s are up, down, and 
strange quark fields; and
\beq
\label{3}
         G^{\mu\nu}&=& \sum_{a=1}^8 \frac{\lambda_a}{2} G_a^{\mu\nu}
\; ,
\eeq
with $\lambda_a$ the SU(3) generator ($Tr[\lambda_a \lambda_b]
= 2 \delta_{ab}$).

   First we shall use the method of QCD Sum Rules to estimate the mass of the
lowest strange hybrid baryon and see if it matches the mass of the 
$\Lambda(1405)$. Then we shall use a current which is an admixture of a hybrid
and a normal strange $IJ^P=0(1/2)^-$ baryon. 

\subsection{Review of QCD Sum Rules for a strange hybrid baryon} 

The correlator for use in the QCD sum rule method for a strange hybrid 
baryon with $IJ^P = 0[\frac{1}{2}]^-$ 
is 
\beq
\label{4}
   \Pi^H(x) &=& <0|T[J_{H}(x) J_{H}(0)]|0> \; ,
\eeq

The QCD sum rule method is based on equating a dispersion relation of the
 correlator, called the left hand side, to an
operator product expansion (OPE) of the correlator, called the right hand
side. For our hybrid hypothesis the dispersion relation is
\beq
\label{5}
 \Pi(q)^H_{\rm{lhs}} &=&  \frac{\rm{Im}\Pi^H(M_H)}
{\pi(M_H^2-q^2)}+\int_{s_0}^\infty ds \frac{\rm{Im}\Pi^H(s)}
{\pi(s-q^2)}
\eeq
where $M_H$ is the mass of the state (assuming zero width)
and $s_0$ is the start of the continuum--a parameter to be determined.
The imaginary part of $\Pi^H(s)$, with the term for the state we are
seeking shown as a pole (corresponding to a $\delta(s-M_H^2)$ term in 
$\rm{Im}\Pi$), and the 
higher-lying states produced by $J_H$ shown as the continuum, is 
illustrated in Fig. 1
\begin{figure}[ht]
\begin{center}
\epsfig{file=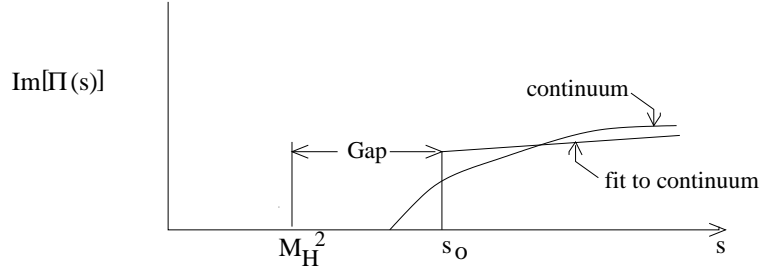,height=4cm,width=10cm}
\caption{QCD sum rule study of a state H with mass M$_H$ (no width)}
\label{Fig. 1}
\end{center}
\end{figure}

Next $ \Pi^H(q)$ is evaluated by an operator product expansion
(O.P.E.), giving the right-hand side (rhs) of the sum rule
\beq
\label{6}
  \Pi(q)_{\rm{rhs}}^A &=& \sum_k c_k(q) \langle 0|{\cal O}_k|0\rangle
 \; ,
\eeq
where $c_k(q)$ are the Wilson coefficients and $\langle 0|{\cal O}_k|0\rangle$
are gauge invariant operators constructed from quark and gluon fields,
with increasing $k$ corresponding to increasing dimension of ${\cal O}_k$.
It is important to note that the Wilson coefficients, $c_k(q)$ obey
renormalization group equations\cite{rry85}

  After a Borel transform, ${\mathcal B}$, in which the q variable is 
replaced by the Borel mass, $M_B$, the final QCD sum rule has the form
\beq
\label{7}
   {\mathcal B} \Pi^H(q)(LHS) &=& {\mathcal B} \Pi_A(q)(RHS)\; .
\eeq

For a succesful solution the value of $M_H(M_B)$ should have a minimum
near the value of $M_B$, and should not depend very much on $M_B$. Also, 
the value of $s_o$ should be approximately the magnitude of the square of the
mass of the next higher excited state.

\subsection{QCD Sum Rule for a Strange Hybrid $IJ^P= 0(1/2)^-$ baryon.}

  It is useful to divide the correlator into a vector and scalar part,
$\Pi^H_V$ and $\Pi^H_S$:
\beq
\label{8}
  \Pi^H(p) &=& \int d^4x e^{1p\cdot x} \Pi^H(x) = \Pi^H_V(p) \not\!{p}
+ \Pi^H_S(p)
\eeq

 For our present problem we use diagrams up through dimension six. 
The diagrams needed are shown in Fig.2.

   The lowest dimension term, shown as 1 in Fig. 2, is
\beq
\label{9}
 \Pi^H_1(x) &=& \frac{1}{2} \sum_{a,b,c,d} (\epsilon^{abc})^2 g^2 
[Tr[\gamma_\mu S_s(x) \gamma_\nu S_u(-x)] S_d(x)+  \\
&&Tr[\gamma_\mu S_s(x) \gamma_\nu S_d(-x)] S_u(x)-Tr[S_u(-x)
\gamma_\mu S_s(x) \gamma_\nu S_d(-x)] \nonumber \\
&&-Tr[S_d(-x) \gamma_\mu S_s(x) \gamma_\nu S_u(x)]] 
 Tr[\frac{1}{4}\lambda^d \lambda^e] \gamma^\sigma \gamma^\lambda 
<G^d_{\mu \sigma}(x) G^e_{\nu \lambda}(0)> \nonumber \; ,
\eeq
with $S_f(x)$ the f-flavor quark propagator. Neglecting the quark masses
for the u and d quarks, the quark propagators are

\beq
\label{10}
       S_u(x) &=&  S_d(x) = S(x)=\frac{i}{2 \pi^2} \frac{\not\!{x}}{x^4}
\nonumber \\
         S(p) &=& \frac{\not\!{p}}{p^2} \\
         S_s(p) &=& S(p) + \frac{m_s}{p^2} \nonumber \; ,
\eeq
where $m_s$ is the strange quark mass, and we assume $1/(p^2 -m_s^2) 
\simeq 1/p^2$ for the values of $p^2$ which are relavent for the calculation 
of the mass of the hybrid baryon.     

  Thus for the vector part of $ \Pi^H_1(p)$ one must evaluate
\beq
\label{11}
   \Pi^H_{1V}(p) &=&  \int d^4x e^{1p\cdot x} \sum_{a,b,c,d}(\epsilon^{abc})^2
 g^2 Tr[\gamma_\mu S(x) 
\gamma_\nu S(-x)]  \\
 && Tr[\frac{1}{4}\lambda^d \lambda^e] \gamma^\sigma \gamma^\lambda 
<G^d_{\mu \sigma}(x) G^e_{\nu \lambda}(0)> S(x) \nonumber\; . 
\eeq

We do not need the scalar part, $\Pi^H_{1S}(p)$, as is explained below. 

In order to complete the calculation of $\Pi^H_{1V}$  one  needs the 
following:
\beq
\label{12}
 && Tr[\frac{1}{4}\lambda^d \lambda^a] = \frac{1}{2}\delta(d,a) \; ,  \\
 &&<G_{\mu \sigma}(x)G_{\nu \lambda}(0)>= 
\frac{1}{2 \pi^2 x^4} [g_{\sigma \lambda}(g_{\mu \nu} \nonumber \\
  && -\frac{4 x_\mu x_\nu}{x^2})+(\sigma, \lambda) \leftrightarrow (\mu, \nu)
 -\sigma \leftrightarrow \mu -\lambda \leftrightarrow \nu] \; , \nonumber
\eeq

  After a rather complex calculation, using $g^2=4\pi \alpha_s\simeq 4 \pi$,
 we find that

\beq
\label{13}
    \Pi^H_1(p) &=& \frac{i}{5 \cdot 3 \cdot 2^{14} \cdot \pi^5} \not\!{p} p^8 
ln(-p^2) \; 
\eeq

  The next term in the OPE, shown as 2 in Fig. 2, contains the gluon 
condensate. The basic difference from diagram 1 is\cite{rry85}
\beq
\label{14}
    <G^e_{\mu \sigma}(x)G^e_{\nu \lambda}(0)>&\Rightarrow&  
<G^e_{\mu \sigma}(0)G^e_{\nu \lambda}(0)> \\
       <G^e_{\mu \sigma}(0)G^e_{\nu \lambda}(0)>&=& 
\frac{ (g_{\mu \nu} g_{\sigma \lambda}-g_{\mu \lambda} g_{\nu \sigma}) <G^2>}
{{2^5 \cdot 3}} \nonumber \; ,
\eeq

\clearpage

\begin{figure}[ht]
\begin{center}
\epsfig{file=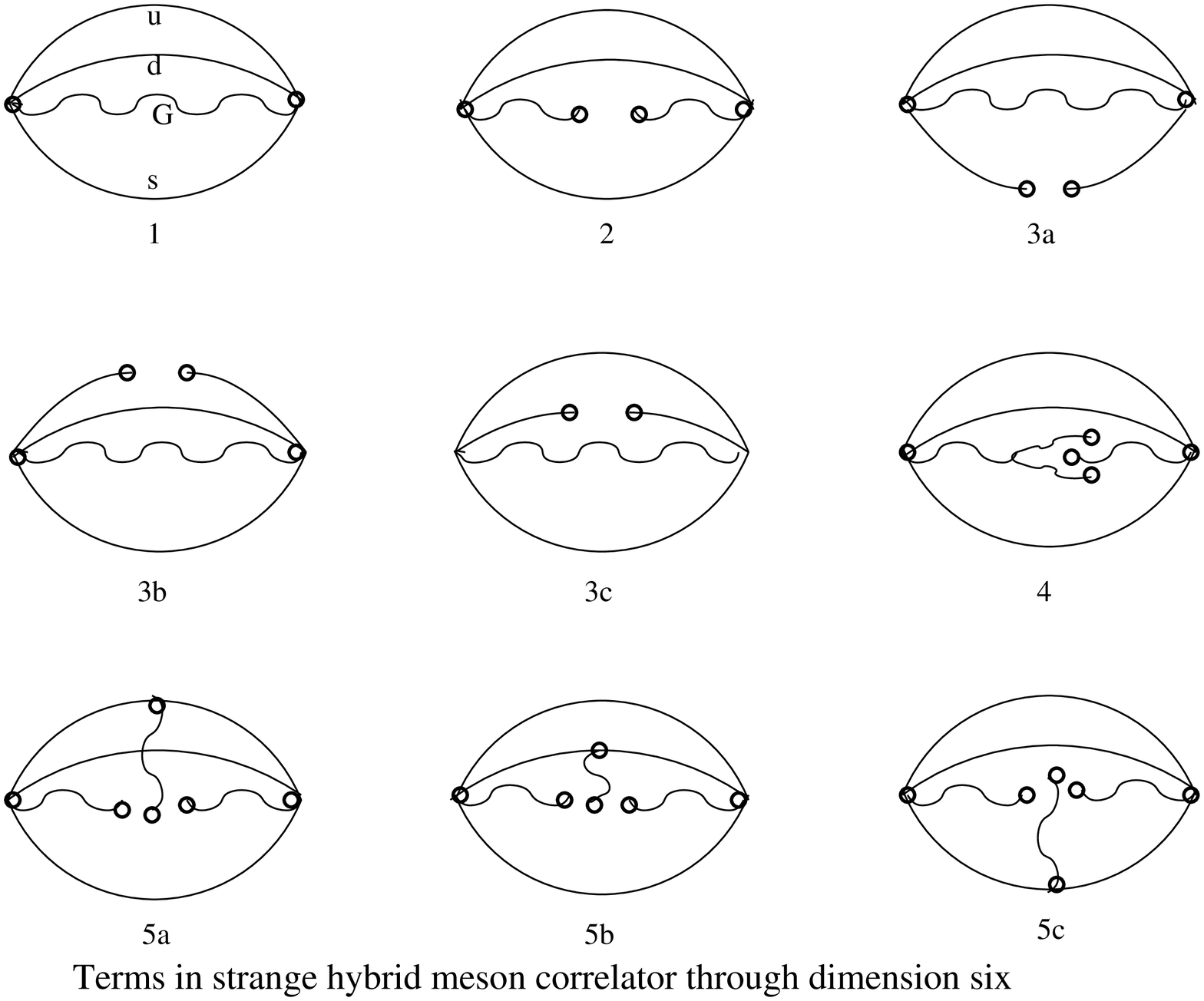,height=8cm,width=12cm}
\caption{}
\label{Fig.2}
\end{center}
\end{figure}

where $<G^2>$ is the gluon condensate. This term of the correlator has only
a vector part, and in momentum space is
\beq
\label{15}
   \Pi^H_2(p)&=& \frac{i <G^2>}{2^{12} \cdot  \pi^3} \not\!{p} p^4
ln(-p^2) \; .
\eeq

  The quark condensate terms are shown as diagrams 3a, 3b and 3c, Fig. 2.
Processes 3b and 3c vanish, while 3a is obtained from process 1 with the 
replacement (see Eqs.({\ref{9},{\ref{10}))
\beq
\label{16}
           S_s(x) &\Rightarrow& \frac{<\bar{s}s>}{12}\; ,
\eeq
where $<\bar{s}s>$ is the strange quark condensate. Therefore we find that
processes 3 have only a scalar part, and in momentum space is
\beq
\label{17}
    \Pi^H_{3S}(p) &=& i \frac{< \bar{s}s>}{2^9 \cdot 3 \pi^3} p^6 ln(-p^2)
 \; .
\eeq

   For process 4, Fig. 2, one makes use of the OPE for the gluon 
field\cite{rry85}:
\beq
\label{18}
      G^e_{\mu \sigma}(x) &\simeq&  G^e_{\mu \sigma}(0) \\
&& +\frac{gx^\beta x^\delta}{4} f_{ebc} G^b_{\mu \beta}(0) 
G^c_{\sigma \delta}(0)+... \nonumber \; ,
\eeq
where the first term gives the gluon condensate in process 2.

To obtain the correlator for process 4 one needs the expression for the
three-gluon vacuum expectation value:
\beq
\label{19}
  && <f_{ebc} G^e_{\nu \lambda}(0) G^b_{\mu \beta}(0) G^c_{\sigma \delta}(0)>
 = \frac{<f_{ebc} G^3>}{24} \nonumber \\
&& [g_{\nu \delta} g_{\mu \lambda} g_{\beta \sigma}+ {\rm \;seven\;similar\;
terms}]
 \; ,
\eeq
with $<f_{ebc} G^3>$ the six-dimensional three-gluon condensate.

After a complicated calculation one finds for the vector correlator for 
process 4
\beq
\label{20}
       \Pi^H_{4V}(p)&=& -\frac{i <g^3 f G^3>}{2^7 \cdot 3 \pi^4} \not\!{p}
p^2 ln(-p^2) \; .
\eeq

  The final dimension six processes are shown in diagrams 5a, 5b, and 5c,
FIG. 2. One can show that the sum of the two terms with gluon fields
coupled to the u and d quarks, processes 5a and 5b, cancel; i.e.,
$\Pi_{5a} + \Pi_{5b} =0$. For the 5c process, with a gluon field coupled to
the strange quark, the calculation is similar to the calculation for process 2,
with a gluon condensate, with the propagator for the s quark replaced by
the s quark propagator with an external gluon field, as shown in process 5c of
Fig. 2. Thus the expression for the s quark propagator becomes
\beq
\label{21}
         S_s(x) &\Rightarrow& \frac{i}{2 \pi^2} \frac{g}{2^4}
\frac{\not\!{x}\sigma^{\rho \delta} +\sigma^{\rho \delta}\not\!{x}}
{x^2} G_{\rho \delta}+{\rm \;scalar\;part} \; .
\eeq

After a lengthy calculation one finds for the vector part of process 5
in momentum space
\beq
\label{22}
      \Pi^H_{5V}(p)&=& \frac{i <g^3 f G^3>}{2^8 \cdot 3 \cdot \pi^4} \not\!{p}
p^2 ln(-p^2) \; .
\eeq

  The only nonvanishing mixed quark condensate only contributes to the
scalar part of the correlator, like process 3, which does not enter in our
estimate of the strange hybrid mass (see below), and is therefore not
considered. The final dimension six diagram is the four-quark condensate,
proportional to $< \bar{u}u>< \bar{d}d>$. Using $< \bar{q}q>\simeq 0.014 GeV^3$
one finds that this diagram is less than 10 \% of the leading diagram, and
is neglected.

In summary, separating the hybrid correlator into a vector and scalar part,
with the vector part
\beq
\label{23}
    \Pi^H (p) &\equiv& \Pi^H_V(p) \not\!{p} + \Pi^H_S(p) \; ,
\eeq
\beq
\label{24}
    \Pi^H_V(p)&=&i \frac{ln(-p^2)}{3 \cdot 2^7 \cdot \pi^4} 
[ \frac{g^2 p^8}{5 \cdot 2^7 \cdot \pi} +\frac{3 \cdot \pi g^2<G^2>p^4}{2^5}
\nonumber \\
 && - <g^3 f G^3> p^2 (1.-0.5)] \; .
\eeq

We use the standard values for the gluon condensates, $<\alpha_s G^2>$ 
= 0.0377 GeV$^2$ and $<g^3 fG^3>$ = 0.0422 GeV$^6$, with $g^2/4\pi
\simeq 1.0$. Note that in Refs\cite{kpr09,lsk09} only the scalar part of the
correlator is used to estimate the hybrid meson mass.
The vector correlator is more reliable for the present calclculation, as the
scalar part mainly has higher dimensional contributions,
and we only use the vector correlator
to estimate the strange hybrid mass in our present work. In many applications
of the QCD Sum Rule Method one obtains the mass of the pole term by taking
the ratio of the vector to scalar correlator, or similar techniques. This
removes unknown constants. Another standard technique is to use either the
vector or scalar component and take the ratio to the sum rule to its derivative
with respect to $1/M_M^2$, which is the method we use in the present work.

   First note that the Borel transform has the following properties:
\beq
\label{25}
     {\mathcal B} \frac{1}{p^2-s} &=& -e^{-s/M_B^2} \nonumber \\
      {\mathcal B}[(p^2)^k ln(-p^2)] &=& -k (M_B^2)^{k+1} \; .
\eeq

From this we obtain the left and right hand sides of the correlator as
a function of the Borel mass (in units of GeV):
\beq
\label{26}
 && \Pi^H_V(M_B)_{rhs}= \frac{i}{3 \cdot 2^8 \cdot \pi^4}[0.3(M_B^2)^5 
 +0.56 (M_B^2)^3 \nonumber \\
 &&- 0.0211 (M_B^2)^2] \; {\rm and} \\
 && \Pi^H_V(M_B)_{lhs}= F e^{-M_H^2/M_B^2} + e^{-s0/M_B^2}(K_0 + K_1 M_B^2 
\nonumber \\
 &&+K_2 M_B^4 +K_3 M_B^6+K_4 M_B^8 ) \nonumber \; ,
\eeq
where $F$ is the amplitude of the pole term, the $K_i$ are chosen to fit
the continuum, and $s0$ is the value of $s$ at the start of the fit to
the continuum.

   Evaluating the sum rule by combining it with a derivative 
$\partial/\partial(1/M_B^2)$ of the sum rule (see, e.g., Ref\cite{lsk09}), 
and redefining the constants $c_i=-3 \cdot 2^8 \cdot \pi^4/i\times K_i$,
one obtains the expression for $(M_H)^2$ as a function of $(M_B)^2$
\beq
\label{27}
  M_H^2 &=& (e^{-s_0/M_B^2}[ s_0(c_0 +c_1 M_B^2 +c_2 M_B^4 + c_3 M_B^6
\nonumber \\
  &&+c_4 M_B^8) +c_1 M_B^4 + 2 c_2 M_B^6 + 3 c_3 M_B^8 + 4 c_4 M_B^{10} ] 
\nonumber \\
&& +1.5 M_B^{12}  + 1.68 M_B^8 -0.0422 M_B^6 ) \times \nonumber \\
 &&  (e^{-s_0/M_B^2}(c_0 +c_1 M_B^2 +c_2 M_B^4 + c_3 M_B^6+c_4 M_B^8)
 \nonumber \\
&&+0.3 M_B^{10} +0.56 M_B^6 -0.0211 M_B^4 )^{-1} \; .
\eeq

The solution to the sum rule is shown in FIG 3. Following the standard
QCD Sum Rule procedure, the parameters $s_0,c_i$ are chosen so the value
of $M_H^2$ has a minimum and has a very weak dependence on $M_B^2$ in
the region of the minimum. The solution shown in FIG 3 is obtained with
$s_0$=2.5 GeV$^2$ and $c_0=230.,\;c_1=-38.,\;c_2=8.1,\;c_3=-7.6,\;
c_4=-.088$, with appropriate powers of GeV.
\begin{figure}[ht]
\begin{center}
\epsfig{file=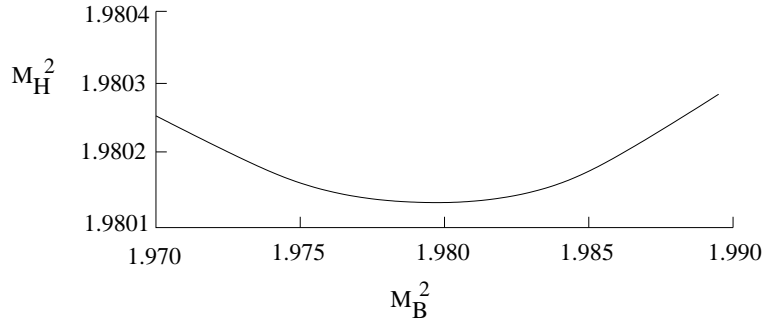,height=4cm,width=10cm}
\caption{Solution to QCD Sum Rule. The units for the masses are GeV}
\label{Fig.3}
\end{center}
\end{figure} 

With $s_0$=2.5 GeV$^2$ and the values of $c_i$ given above, the pole 
term approximately 40\% larger than the continuum. This satisfies the 
criteria for a good solution. Using the criteria that have been used for 
decades we estimate that the error in our calculation of the mass is about 
15 per cent.
As one can see from the figure the mass of the lowest strange hybrid with
$IJ^P=0(1/2)^-$ is 1407 MeV, which is approximately the mass of the 
$\Lambda(1405)$. We emphasize that by including all graphs through dimension
six, we ensure the convergence of the operator product expansion. Note
that the dimension six contributions to the correlator after the Borel
transformation are very small.

In thre next section we consider a mixed hybrid-three quark model for the
$IJ^P=0(1/2)^-$ $\Lambda(1405)$.

\section{$\Lambda(1405)$ as a mixed hybrid/3-quark strange baryon}

   In our study of heavy quark mesons\cite{kpr09}, no hybrid charmonium
solution was found; and in further studies of charmonium and upsilon states
it was shown\cite{lsk09} that the $\Psi'(2S)$ and $\Upsilon(3S)$ states
are a mixed hybrid and normal meson. In the study  of the $\Lambda(1405)$ 
by Nakamura $et\;al$\cite{oka08} QCD Sum Rules were used with a mixed
three-quark ($J_3$) and pentaquark ($J_5$) current. Although a stable 
solution was not found, this study showed that the 3-quark component was small.
   We now investigate the possibility that there is a QCD Sum Rule
solution with a correlator given by a mixed hybrid and 3-quark current:
\beq
\label{28}
              J(x) &=& b J_H(x) + \sqrt{1-b^2} J_3 \; ,
\eeq
and the correlator
\beq
\label{29}
                \Pi_{H-3}(x) &=& <0|T[J(x)J(0)|0> \\
     &=& b^2 \Pi_H(x) +(1-b^2)\Pi_3(x) +2b\sqrt{1-b^2} \Pi_{H3}(x)
\nonumber \\
     \Pi_H(x)&=& <0|T[J(x)_HJ_H(0)|0> \nonumber \\
      \Pi_3(x)&=& <0|T[J(x)_3J_3(0)|0> \nonumber \\
       \Pi_{H3}(x)&=& <0|T[J(x)_HJ_]3(0)|0> \nonumber \; ,
\eeq
Where $J(x)_H$ is the hybrid current, defined in Eq(\ref{2}) and $J_3$
is the three-quark current.
\clearpage

\begin{figure}[ht]
\begin{center}
\epsfig{file=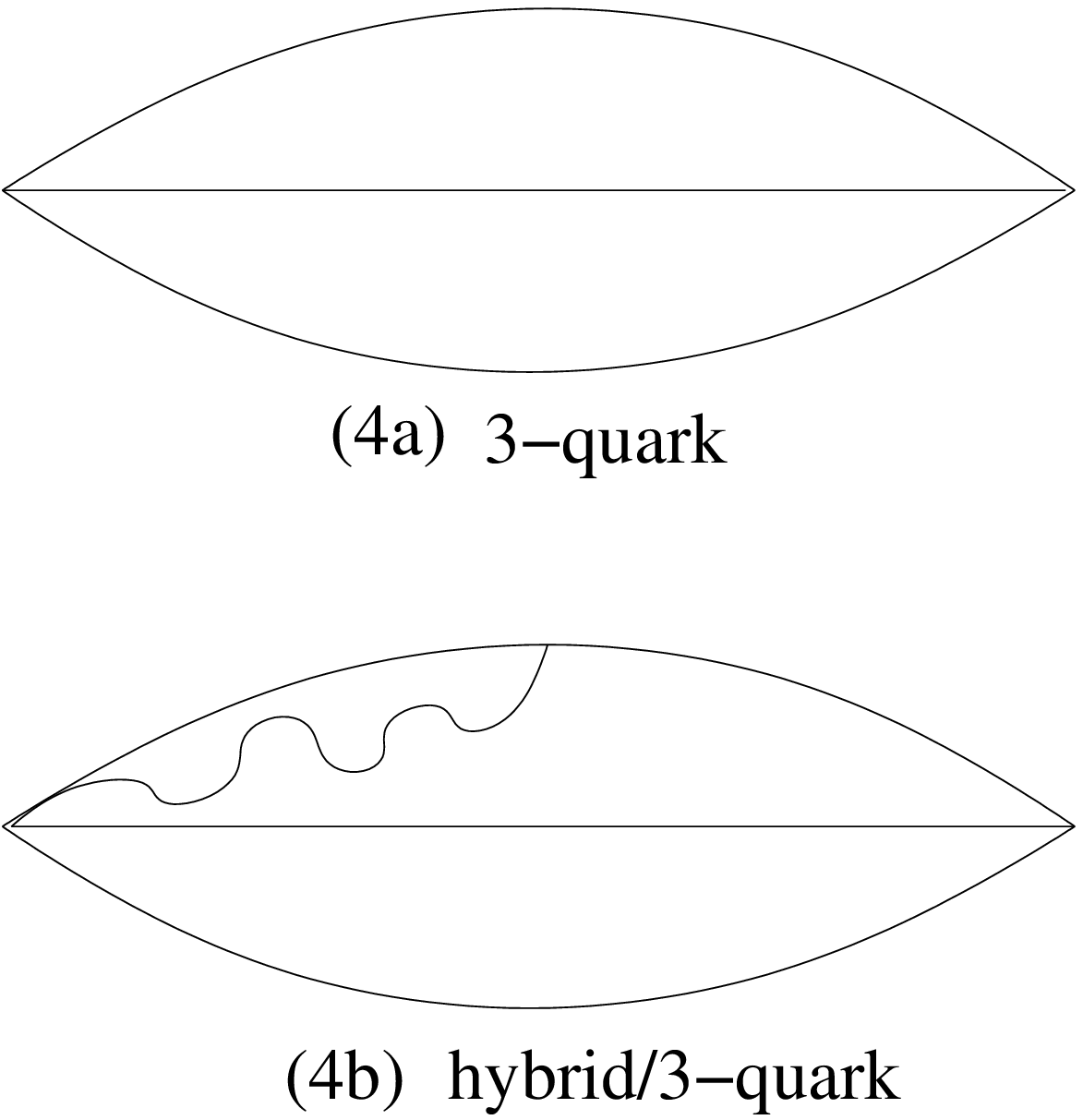,height=6cm,width=10cm}
\caption{Diagrams for the 3-quark and coupled hybrid/3quark correlators}
\label{Fig.4}
\end{center}
\end{figure} 

 We could use the same $J_3$ for a strange $IJ^P=0(1/2)^-$ state as in 
Ref\cite{oka08}, but since it has different dimensions than $J_H$ for our 
calculation or $J_5$ with a mixed $J_3$ and $J_5$, current, the Sum Rule 
method cannot be used without a method of normalizing the two components 
of the current.  In Ref\cite{oka08} a model for the wave functions was used 
to obtain the dimensional normalization. In our present work we wish to
use QCD direclty. Using the method derived for the mixed 
heavy-quark/hybrid-heavy-quark QCD Sum Rule calculation\cite{lsk09}
we obtain a crucial relationship between $\Pi_3(x)$ and $\Pi_{H3}(x)$ 
correlators, which are shown in Fig 4. 

By using the external field method, as in Ref\cite{lsk09} one can show that
\beq
\label{30}
            \Pi_{H3}(x) &\simeq& \pi^2 \Pi_3(x) \; .
\eeq

Noting that $\Pi_H(x)$ and  $\Pi_{H3}(x)$ have the same dimension, we use
for the mixed hybrid/3-quark strange $IJ^P=0(1/2)^-$ correlator
\beq
\label{31}
        \Pi_{H-3}(x) &=&  b^2 \Pi_H(x) +(\frac{(1-b^2)}{\pi^2}
+2 \sqrt{1-b^2}) \Pi_{H3}(x) \; .
\eeq

  Note that there are three diagrams for $\Pi_{H3}$, shown in Fig.4b.
The gluon from the hybrid vertex can couple to the u, d, or s quark. The
first two cancel, and $\Pi_{H3}$ is given by the coupling to the strange 
quark:
\beq
\label{32}
     \Pi_{H3}(x) &=& -i\frac{g^2}{3} Tr[S(x) \gamma_\nu S(x)\gamma_\mu]
\gamma^\mu [\sigma_{\kappa \delta},S(x)]_{+} \gamma_\lambda 
 Tr[G^{\nu \lambda}(0) G^{\kappa \delta}(0)] \; .
\eeq
All of the quantities in Eq(\ref{32}) have been defined above except 
the anticommutator $[A,B]_{+}=AB+BA$, and $\sigma_{\kappa \delta}=
i(\gamma_\kappa \gamma_\delta-g_{\kappa \delta})$.

After a somewhat complicated calculation and a Fourier transform, we find
that with the neglect of quark masses, as in the calculation of $\Pi_{H}$,
the only term with dimension 6 or less has the form $g^2 <G^2> p^4 ln(-p^2)$,
proportional to the second term in $\Pi^H_V(p)$ shown in Eq(\ref{26}). 
Therefore, after a Borel transform $\Pi_{H3}(M_B)$ will be proportional to
$(M_B^2)^3$ as is the second term in $\Pi^H_V(M_B)$, Eq(\ref{26}). One finds 
(in comparison to the second term in $\Pi^H_V$, $\Pi^H_2$) that 
\beq
\label{33}
       \Pi_{H3}(M_B)&=& \Pi^H_2(M_B) \times (0.2) \; .
\eeq

   Thus we find that the contribution to the mixed hybrid/3-quark
correlator from the 3-quark component is approximately 20 \% of the 
second-order term in the hybrid correlator for any value of b. Noting
that terms of this magnitude have been dropped in obtaining the hybrid
correlator, as discussed in Section 2, we conclude that our calculaion of 
the $\Lambda(1405)$ as a hybrid, discussed in the previous section, is 
unchanged within errors of the method of QCD Sum rules. Our result is 
analogous to that in Ref\cite{oka08} in which it was found that the 3q 
component was very small in their 3q/5q model.

   It is interesting to compare our present calculation of the $\Lambda(1405)$
as a mixed mixed hybrid/3-quark state to our previous calculation of the
Roper as a mixed hybrid/3-quark state, with the 3-quark component being
nucleon-like. In that calculation\cite{kl95} it was found that the 3-quark
component of the Roper was small, and that the Roper is essentially a pure
hybrid baryon, similar to our conclusion about the nature of the
$\Lambda(1405)$.

We  now discuss possible experimental tests of the hybrid nature of the 
$\Lambda(1405)$.

\section{Experimental tests of the $\Lambda(1405)$ as a hybrid}

   The only decay mode of the $\Lambda(1405)$ that has been measured is
the decay into a $\Sigma+\pi$. It would be interesting to calculate the
$\pi^+$ vs $\pi^-$ decay\cite{oset09}, for which we need to extract the
equivalent to the $\Lambda(1405)$ wave function. This is not provided
by the QCD Sum Rule, and will be the subject of a future investigation.
Moreover, we shall investigate the photoproduction of the $\Lambda(1405)$,
and expect very different results than those found using a bag 
model\cite{barnes83b}

   Polarization could also be an important test of the nature of the 
$\Lambda(1405)$. Polarization of $\Lambda$ in diffractive pp collisions was
measured some years ago\cite{r608}. See Ref.\cite{zb00} for a theoretical 
study using a standard model for the $\Lambda(1405)$ baryon, which is not
consistent with the data, and references to earlier publications.
Note that the polarization of the $J/\Psi$, which was measured by the CDF 
Collaboration\cite{CDF}, was shown\cite{bkl00} to disagree with the standard 
nonrelativistic QCD factorization method, and might be a test of the nature 
of charmonium states. Similarly, the polarization of the 
$\Lambda(1405)$ could be used to test the hybrid nature of that state.
Measurements of the photoproduced $\Lambda(1405)$ are in progress\cite{rs09}.
One mechanism for polarization of the hyrid $\Lambda(1405)$ is shown in
Figs. 5 and 6.
  
   The figures show a photon producing a $s \bar{s}$ pair, which leads to
the process $\gamma + p \rightarrow \Lambda (1405) + K^+ $, as in the usual
photoproduction process. For our case, however the $s$ quark (or $\bar{s}$
quark) emits an octet gluon before combining with  $u$ and $d$ quarks
to form a hybrid $\Lambda(1405)$ The vector gluon would then lead to quite
different polarzation than in the same process without the gluon, as in
a standard $u, d, s$ strange baryon. The study of such processes for the 
polarization of the $\Lambda(1405)$ created via photoproduction, shown in
Fig. 6, is part of our future research.
\clearpage

\begin{figure}[ht]
\begin{center}
\epsfig{file=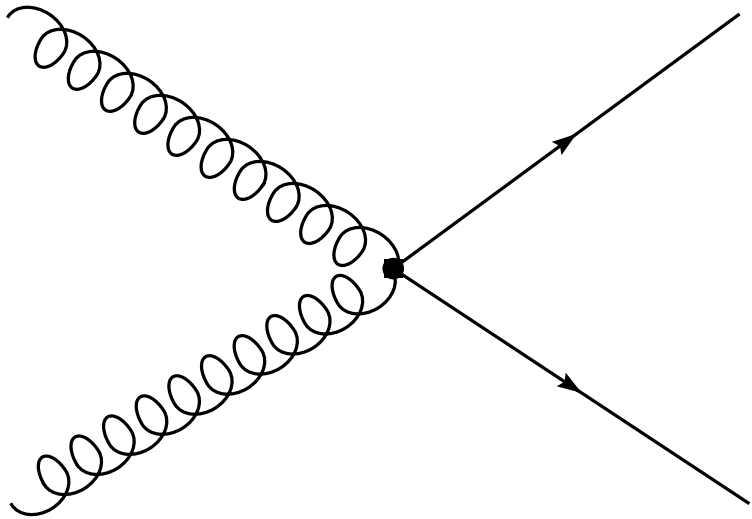,height=2.0cm,width=10cm}
\caption{Scalar glueball-meson coupling theorem}
\label{Fig.5}
\end{center}
\end{figure}

\begin{figure}[ht]
\begin{center}
\epsfig{file=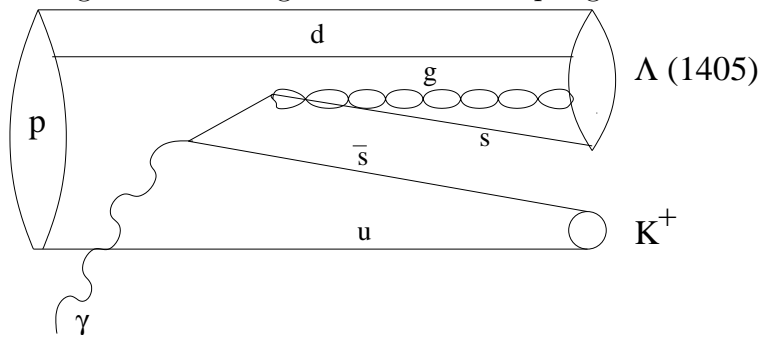,height=3cm,width=10cm}
\caption{Polarization of a photoproduced hybrid $\Lambda(1405)$}
\label{Fig.6}
\end{center}
\end{figure}
\vspace{2cm}

\begin{figure}[ht]
\begin{center}
\epsfig{file=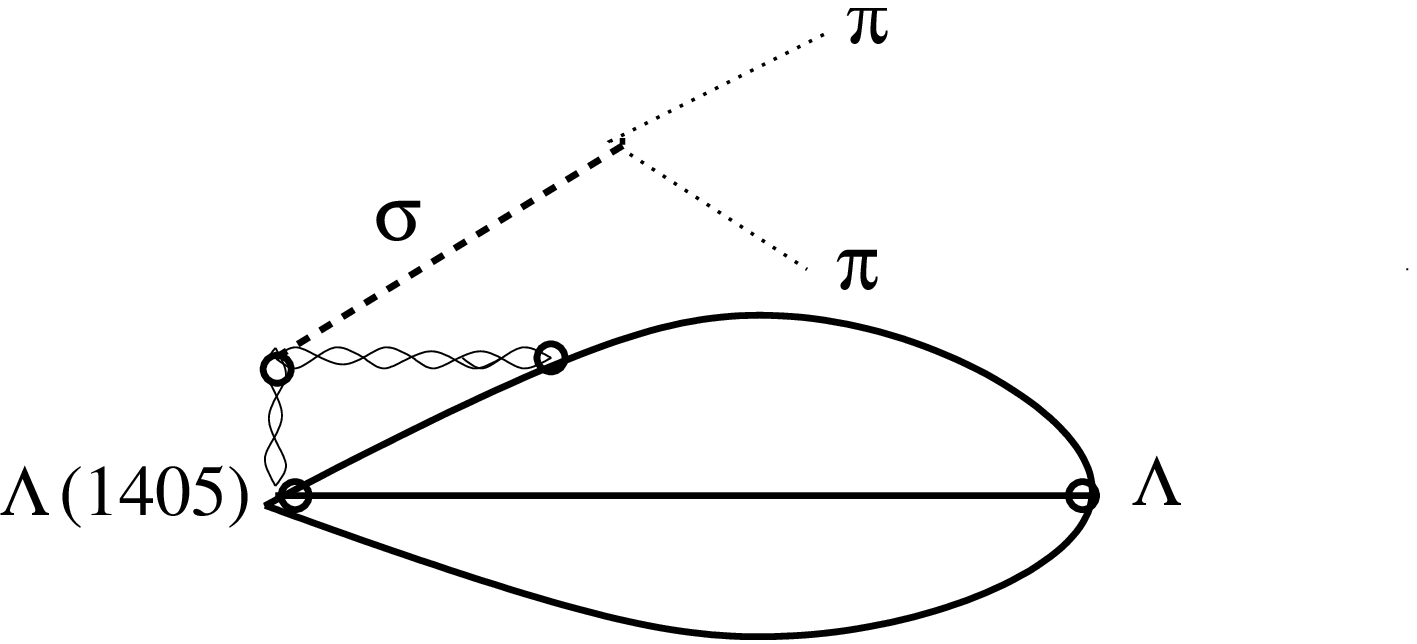,height=3.0cm,width=10cm}
\caption{Sigma decay of a hybrid $\Lambda$(1405)}
\label{Fig.7}
\end{center}
\end{figure}

\clearpage

  An important test of hybrids is $\sigma$ decay. The $\sigma$ is a broad
singlet $\pi-\pi$ resonance at about 600 MeV. Using a theorem on the
coupling of glue to a scalar meson\cite{nov}, illustrated in Fig. 5.
$\sigma$ decay of the Roper\cite{kl95} and the $\Lambda(1600)$\cite{lsk04}
were suggested as experimental tests for the hybrid nature of these states.
The mechanism for the creation of a hybrid  is shown in Fig. 6.
This so-called sigma/glueball model, based on the theorem illustrated in
Fig. 5, has predicted the decay of glueballs into sigmas\cite{lsk97}, which
is consistent with experiment\cite{bes97}.
It was used in Ref.\cite{lsk09} to explain why the $\Upsilon(3S)$
emits a $\sigma$ when decaying to the $\Upsilon(1S)$, while the $\Upsilon(4S)$
and $\Upsilon(2S)$ do not. Sigma decay of the $\Lambda(1405)$ is shown in
Fig. 7. Unfortunately, the mass of the $\Lambda(1405)$ is 
just above the threshold for decay into a $\Lambda +2 \pi$, so one could not 
detect the entire sigma resonance, but there should be a large enhancement of 
$2 \pi$ decay of a $\Lambda(1405)$ into a $\sigma+\Lambda$, which could be
measured. This is a subject of future research.

\section{Conclusions}

   Using the QCD Sum Rule method we find that the $\Lambda(1405)$ is 
consistent with being a strange hybrid baryon. Ours is the only calculation 
which has been done to examine the possibility that this state is a hybrid 
that proceeds directly from QCD. Since this method does not provide the wave 
function for the state, we shall use other methods to estimate the decay 
widths and polarization in the future. The sigma (pi-pi resonance) decay is 
an important test of the hybrid nature of a state, so the decay of the 
$\Lambda(1405)$ into a  $\Lambda + 2 \pi$ could provide a test of our theory. 
\vspace{5mm}
 
\large{{\bf Acknowledgments}}
\normalsize
\vspace{1mm}

This work was supported in part by DOE contracts W-7405-ENG-36 and 
DE-FG02-97ER41014, and by the NSF/INT grant 0529828. The authors thank
Professor Reinhard Schumacher for several helpful discussions.


\begin{thebibliography}{99}
\bibitem{efk65}A. Engler, H.E. Fisk, R.W. Kramer, C.M. Meltzer, J.B. Westgard,
T.C. Bacon, D.G. Hill, H.W.K. Hopkins, D.K. Robinson, and E.O. Salant, Phys.
Rev. Lett. {\bf 15}, 224 (1965)
\bibitem{dalitz61}R.H. Dalitz, Rev Mod Phys. {\bf 33},471 (1961)
\bibitem{dwr67} R.H. Dalitz, T.C. Wong, and G. Rajasekaran, Phys. Rev. 
{\bf 153}, 1617 (1967)
\bibitem{oset09} D. Jido, E. Oset and T. Sekihara, arXiv:nucl-th0/0904.3410;
Eur. Phys. J. {\bf A 39}, 81 (2009)
\bibitem{weise08} T. Hyodo, W. Weise, D. Jido, L. Roca and A. Hosaka,
arXiv:nucl-th/0802.2212; Mod. Phys. Lett. {\bf A 23}, 2393 (2008)
\bibitem{oka08}T. Nakamura, J. Sugiyama, N. Ishii, T. Nishikawa and M. Oka,
Phys. Lett {\bf B662}, 132 (2008)
\bibitem{rs09} Kie Moriya and Reinhard Schumacher, arXiv:nucl-ex/0911.2705;
Proceedings of HYP-X, Nucl. Phys. {\bf A 835} (2010)
\bibitem{close88}F.E.. Close, Rep. Prog. Phys. {\bf 51}, 833 (1988) 
\bibitem{kittel00}O. Kittel and G.R. Farrar, arXiv:hep-ph/0010186 (2000)
\bibitem{kittel05}O. Kittel and G.R. Farrar, arXiv:hep-ph/0508150 (2005)
\bibitem{barnes83a}T. Barnes and F.E. Close, Phys. Lett {\bf B123}, 89 (1983)
\bibitem{barnes83b}T. Barnes and F.E. Close, Phys. Lett {\bf B128}, 277 (1983)
\bibitem{balitsky82} I.I Balitsky, D.I. Dyakonov and A.V. Yung, Phys. Lett.
 {\bf B112}, 71 (1982)
\bibitem{govaerts83}J. Govaerts, F. de Viron, D. Gusbin, and J. Weyers, Phys. 
Lett. {\bf B128}, 262 (1983)
\bibitem{guo07} F-K Guo, P-N. Shen, Z-G. Wang, W-H Liang and L.S. Kisslinger,
arXiv:hep-ph/0703062 (20007)
\bibitem{marty91} A.P. Martynenko, Sov. J. Nucl. Phys. {\bf 54}, 488 (1991)
\bibitem{kl95} L.S. Kisslinger and Z. Li, Phys Rev {\bf D 51}, R5986 (1995)
\bibitem{lsk04} L.S. Kisslinger, Phys Rev {\bf D 69}, 054015 (2004)
\bibitem{kpr09}L.S. Kisslinger, D. Parno and S. Riordan, arXiv:0805.1943,
Advances in High Energy Phys. (2009)
\bibitem{lsk09}L.S. Kisslinger, Phys Rev. {\bf D 79}, 114026 (2009)
\bibitem{cdks81}Y. Chung, H.G. Dosch, M Hremer, and D. Schall, Nucl. Phys.
{\bf B 197},55 (1981)
\bibitem{rry85}L.J. Reinders, H. Rubinstein and Ss Yazaki, Phys. Rep. 
{\bf 127}, 1 (1985)
\bibitem{r608} T. Henkes $et\;al.$, R608 Collaboration, Phys. Lett. {\bf B283},
155 (1992)
\bibitem{zb00} L. Zuo-tang and C. Boros, Phys Rev. {\bf D 61}, 117503 (2000)
\bibitem{CDF} CDF Collaboration, T. Affolder $et\;al$, Phys. Rev. Lett.
{\bf 35} 2886 (2000)
\bibitem{bkl00} E. Braaten, B.A. Kniehl and J. Lee, Phys. Rev. {\bf D 62},
094005 (2000)
\bibitem{nov} V.A. Novikov, M.A. Shifman, A.I. Vainstein, V.I. Zakharov, 
Nucl. Phys. {\bf B165} 67 (1980); Nucl. Phys. {\bf B191} 301 (1981)
\bibitem{lsk97} L.S. Kisslinger, J. Gardner, and C. Vanderstraeten, Phys. 
Lett. {\bf B410}, 1 (1997)
\bibitem{bes97}J.Z. Bai, $et\;al$ (BES Collaboration), Phys. Rev. Lett.
{\bf 76}, 3502 (1996)

\end{thebibliography}
\end{document}